\documentclass{article}
\usepackage{graphics}

\begin{document}

\title{On the Phase Transitions of Graph Coloring and Independent Sets}

\author{Valmir C. Barbosa\thanks{Corresponding author
({\tt valmir@cos.ufrj.br}).}\ \ and Rubens G. Ferreira\\
\\
Universidade Federal do Rio de Janeiro\\
Programa de Engenharia de Sistemas e Computa\c c\~ao, COPPE\\
Caixa Postal 68511\\
21941-972 Rio de Janeiro - RJ, Brazil}

\maketitle

\begin{abstract}
We study combinatorial indicators related to the characteristic phase
transitions associated with coloring a graph optimally and finding a maximum
independent set. In particular, we investigate the role of the acyclic
orientations of the graph in the hardness of finding the graph's chromatic
number and independence number. We provide empirical evidence that, along a
sequence of increasingly denser random graphs, the fraction of acyclic
orientations that are ``shortest'' peaks when the chromatic number increases,
and that such maxima tend to coincide with locally easiest instances of the
problem. Similar evidence is provided concerning the ``widest'' acyclic
orientations and the independence number.
\end{abstract}

\section{Introduction}\label{intr}

The class of decision problems that can be solved in nondeterministic
polynomial time, known as the NP class, is central to the theory of
computational complexity. Informally, a decision problem is in NP if its
solution can be checked to be correct in polynomial time. Obtaining that
solution, however, is altogether a different matter: for some problems it
can be done in polynomial time as well, while for others it is as yet unknown
whether a polynomial-time procedure exists. In order to characterize such
seemingly harder problems, it has proven useful to look at the problems that
are ``complete'' for NP (the so-called NP-complete problems), that is, problems
in NP for which the discovery of a polynomial-time procedure to find a solution
would immediately warrant the existence of such a procedure for all the other
problems in NP as well \cite{gj79}.

The NP-complete problems are thus the hardest problems in NP, and are in a sense
essentially equivalent to one another in terms of how hard they are to be solved
\cite{k72}. But it has been known already for many years, both from practical
experience (e.g., \cite{jt96}) and from looking at the minutiae of the structure
of NP \cite{j90}, that some NP-complete problems are in fact harder than others,
and that a given NP-complete problem may have instances that are significantly
harder than other instances of the same problem. Following some initial results
of about one decade ago \cite{ckt91,msl92}, it is now known that, for
NP-complete problems like satisfiability and its derivations
\cite{mzkst99, gs00} and graph coloring \cite{cg01}, sharp phase transitions
with respect to some order parameter exist and are frequently correlated with
the hardness of finding a solution.

We are concerned in this paper with combinatorial optimization problems. Hard
optimization problems have been characterized in much the same way as their
decision-form counterparts. By a minor technicality, however, they are best
termed NP-hard problems (as opposed to NP-complete problems) to indicate only
that they are at least as hard as the problems in NP (but not necessarily one of
them) \cite{gj79}. Invariably, an optimization problem whose decision-form
variant is NP-complete, is NP-hard.

Unlike decision problems, optimization problems are only now beginning to be
looked at in order to explain the relative hardness of their instances, but we
already have some empirical evidence of the presence of similar phase
transitions \cite{sw01}. In the case of graph coloring, for example, the
optimization problem asks for the graph's chromatic number---the least number of
colors needed to assign one color to each node without ever assigning the same
color to neighbors in the graph (i.e., nodes that are connected by an edge).

What is known for this problem comes from considering a sequence of graphs of
increasing density (number of edges per node) and what happens along this
sequence at the points in which the chromatic number increases. It has been
discovered that finding the chromatic number just before these points is
distinctly harder than just after them. Also, sharp peaks in the size of the
so-called backbone of each graph in the sequence are detected just before those
points as well, thus indicating a strong positive correlation between problem
hardness and backbone size. A graph's backbone in this case is the set of node
pairs that are assigned the same color by every coloring that employs a number
of colors given by the graph's chromatic number. So a graph with a large
backbone presents many opportunities for an algorithm that seeks the optimum to
waste time trying to assign two different colors to a node pair that belongs to
the backbone.

Coloring a graph optimally is one of the problems that we treat in this paper.
While the backbone size relates clearly, in an intuitive way, to why larger
backbones tend to imply harder instances of the problem, we feel that it lends
little combinatorial insight into the hardness of those instances, specifically
into how the structure of the graph affects the hardness of coloring its
nodes optimally. Our contribution in the context of this problem, presented
in Section~\ref{coloring}, is to demonstrate that other indicators exist that
relate just as clearly to the appearance of phase transitions related to the
hardness of graph coloring while at the same time carrying what we think is
better combinatorial intuition. The indicators we use come from considering a
graph's chromatic polynomial and its set of acyclic orientations.

The latter brings us to the second problem of interest in this paper, treated
in Section~\ref{mis}, which is the problem of finding an independent set of
maximum cardinality in a graph. An independent set is a set of nodes that
includes no neighbors. The cardinality of a maximum independent set of a graph
is the graph's independence number. Finding this number is also an NP-hard
problem, one that shares with optimal graph coloring a clean combinatorial
interpretation in terms of the graph's acyclic orientations. This problem does
not appear to have already been examined for phase transitions related to the
hardness of its instances.

Throughout the paper, we use $G(n,m)$ (or simply $G$, if $n$ and $m$ can be
inferred from the context) to denote an undirected graph with $n$ nodes and
$m$ edges. All our empirical results are based on fixed sequences of graphs,
each graph with $n$ nodes but increasingly more edges (up to $N=n(n-1)/2$). For
$1\le s\le N$, the $s$th graph in this sequence, denoted by $G_s(n,s)$ or simply
by $G_s$, has $n$ nodes, $s$ edges, and is generated according to the
random-graph model that samples uniformly from the set of all graphs having the
same number of nodes and edges \cite{b01} (equivalently, for $s\ge 1$, $G_{s+1}$
may be regarded as being obtained from $G_s$ by the random addition of a new
edge).

Using one single sequence as the basis of each experiment precludes the
smoothing effect of taking averages over larger ensembles, known to mask the
appearance of very sharp phase transitions for problems like graph coloring
\cite{cg01,sw01}. We are always careful, however, to make sure that the observed
phenomena are also present, qualitatively, in several other sequences.

For small values of $s$, a graph $G_s$ in the sequence $G_1,\ldots,G_N$ is
likely to have isolated nodes (nodes without neighbors). Such nodes do not
affect the graph's chromatic number, and affect its independence number only
trivially (every isolated node is a member of all maximum independent sets).
So the sequence that is actually used in all our experiments is the sequence
$H_1,\ldots,H_N$, where, for $1\le s\le N$, $H_s$ is obtained by stripping
$G_s$ of its isolated nodes.

Before proceeding, we pause momentarily to consider this issue of isolated
nodes more carefully. Let $\nu_s$ denote the number of isolated nodes of $G_s$.
We have $\nu_1=n-2$, while for $s\ge 1$ it is easy to see that
\begin{equation}
\label{diffeq}
\nu_{s+1}=p_0\nu_s+p_1(\nu_s-1)+p_2(\nu_s-2),
\end{equation}
where $p_k$, for $k\in\{0,1,2\}$, is the probability that the addition of the
$s+1$st edge to $G_s$ incorporates $k$ new nodes into $H_s$. These three
probabilities can be assessed easily as fractions of $N-s$ and lead to a
simplification of (\ref{diffeq}) as
\begin{equation}
\nu_{s+1}=\left(1-\frac{n-1}{N-s}\right)\nu_s,
\end{equation}
which is clearly solved by
\begin{equation}
\nu_s=n\prod_{k=0}^{s-1}\left(1-\frac{n-1}{N-k}\right).
\end{equation}
Furthermore, for $s\ll N$ we obtain
\begin{equation}
\label{isolated}
\nu_s\approx n\left(1-\frac{n-1}{N}\right)^s\approx ne^{-2s/n},
\end{equation}
so $\nu_s$ obviously decreases rapidly with $s$.

We now turn to our two main sections. At the end, concluding remarks are given
in Section~\ref{concl}.

\section{Graph coloring}\label{coloring}

\begin{figure}[t]
\centering
\scalebox{0.5}{\includegraphics{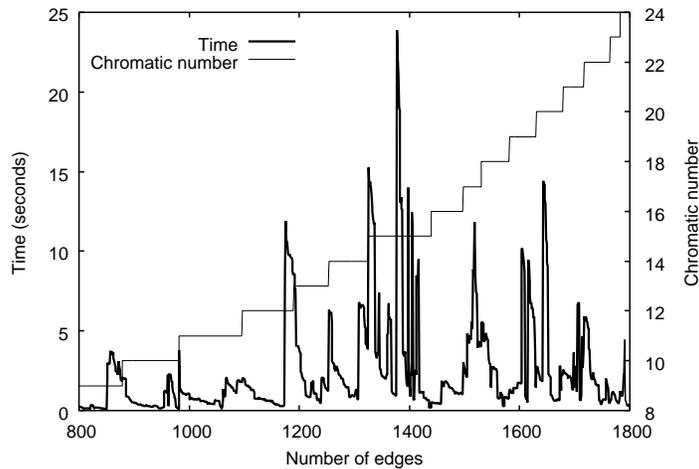}}
\caption{Time to find the chromatic number of some of the graphs in
$H_1,\ldots,H_N$ for $n=65$.}
\vspace{0.2in}
\label{trick}
\end{figure}

Let $\chi(G)$ denote the chromatic number of $G$. We show in Figure~\ref{trick}
two plots for $n=65$, one indicating the time needed to find $\chi(H_s)$ by a
public-domain code \cite{trick.c-url} that is based on the heuristic of
\cite{b79}, the other indicating the evolution of $\chi(H_s)$ as $s$ is
increased. We only show data for a certain range of $s$ values, since coloring
instances outside this interval tend to be relatively trivial, thus requiring
little time for solution. As expected, the chromatic number increases steadily
as the graph gets denser (acquires more edges), and does so increasingly
rapidly. Also, it is often the case that the time needed to find the chromatic
number goes down significantly immediately after each increase in the chromatic
number. In this section, we develop new combinatorial arguments that show that
such sudden transitions are indeed to be expected.

Our study of the hardness of finding $\chi(G)$ starts with an investigation of
how abundant optimal colorings of $G$ are, that is, colorings that require
exactly $\chi(G)$ colors. In order to carry out this investigation, we resort
to the chromatic polynomial of $G$, denoted by $\pi(G,x)$, which gives the
number of distinct ways in which $G$ can be colored by at most $x\ge 0$ colors.

This polynomial has several interesting properties. For example, it is a
degree-$n$ polynomial in $x$, the coefficient of $x^n$ is $1$, and the
coefficient of $x^{n-1}$ is $-m$ \cite{b98}. Also, by definition $\chi(G)$ is
the least value of $x$ for which $\pi(G,x)$ is positive, giving the number of
distinct ways in which $G$ can be colored optimally. So finding $\pi(G,x)$ is
expected to be no easier than finding $\chi(G)$, although the following simple
method can be used for relatively small graphs.

Let $e$ be any edge of $G$, denote by $G\backslash e$ the graph obtained from
$G$ by removing $e$, and by $G/e$ the graph obtained from $G$ by contracting the
end nodes of $e$ into one single node. We see that $G$ can be colored in as
many distinct ways as $G\backslash e$ can, except for those colorings of
$G\backslash e$ that assign to the end nodes of $e$ the same color. But these
are precisely the colorings of $G/e$, so we get
\begin{equation}
\label{recursion1}
\pi(G,x)=\pi(G\backslash e,x)-\pi(G/e,x).
\end{equation}
Clearly, (\ref{recursion1}) defines a simple recursion for calculating
$\pi(G,x)$ that stops either at graphs that only contain isolated nodes or
at graphs that are completely connected. If $k$ is the number of nodes in
either case, then the former graphs admit $x^k$ distinct colorings and the
latter $x!/(x-k)!$ distinct colorings, so the polynomial can be calculated
easily at the bases of the recursion and upward from them.

\begin{figure}[t]
\centering
\scalebox{0.5}{\includegraphics{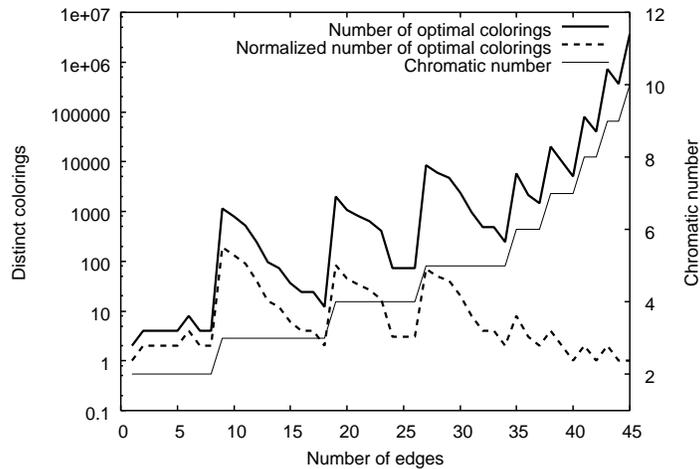}}
\caption{Number of distinct optimal colorings of each graph in
$H_1,\ldots,H_N$ for $n=10$.}
\vspace{0.2in}
\label{poly10}
\end{figure}

In Figure~\ref{poly10}, three plots are given for $n=10$: one depicts the
continual increase of the chromatic number as the number of edges is increased,
while another shows the number of distinct optimal colorings for each graph
$H_s$, that is, $\pi(H_s,\chi(H_s))$ (the third plot is discussed shortly in
what follows). For each graph, the chromatic polynomial has been computed using
public-domain code \cite{chromatic.c-url} based on the recursion of
(\ref{recursion1}). This computation quickly exhausts processing and memory
resources as the numbers of nodes and edges increase---thence the reason why we
present data for $n=10$ only. So we must always bear in mind that our ability to
draw general conclusions may be impaired.

As the plots of Figure~\ref{poly10} indicate, the number of distinct optimal
colorings increases dramatically at each step in the chromatic number, and after
that decreases more or less steadily until immediately before the next step. In
absolute terms, then, at each increase in the chromatic number optimal colorings
become strictly more abundant. If the same could be shown to hold in relative
terms as well (i.e., if the fraction of optimal colorings relative to some
larger universe could be shown to undergo the same transition), then we might be
able to continue and investigate in more detail how this relates to the time it
takes to color the graphs optimally. However, we find that it is as yet unclear
how to characterize such a larger universe.

Even so, it is worth examining the matter further, because at least in part the
sudden jumps in the number of optimal colorings are really the product of well
understood combinatorial growth at work. For example, let us examine what
happens when the chromatic number increases from $k$ to $k+1$ along the sequence
of graphs. Suppose nodes $i$ and $j$ are the nodes that, when connected to each
other by an edge, cause the graph's chromatic number to increase. Clearly, each
optimal coloring before the increase (i.e., with $k$ colors) yields at least
$k+1$ distinct optimal colorings after the increase (i.e., with $k+1$ colors):
viewing colors as positive integers, first choose one of $i$ or $j$ and assign
to it color $k+1$, while the remaining nodes all retain their previous colors;
then assign color $k+1$ to all nodes that had color $k$ and this color to the
one of $i$ or $j$ that was selected previously; then proceed likewise until this
same node has been assigned all colors ($k+1$ down through $1$).

While this may all seem like obvious combinatorial growth at the points where
the chromatic number changes, we remark that such an effect may also be present
in the behavior of the backbones commonly used to characterize the phase
transitions of graph coloring. In the setting that we just examined, nodes $i$
and $j$ clearly constitute one of the node pairs of the backbone---or else the
addition of an edge between them would not cause the chromatic number to
increase, because there would be at least one optimal coloring with $k$ colors
that would assign different colors to them. So let us consider the probability
that randomly chosen nodes $i$ and $j$ constitute a backbone pair.

We do so by first conceding, just for the sake of the argument, that nodes are
uniformly distributed among the colors in all optimal colorings. In this case,
the probability that $i$ and $j$ have the same color in all optimal colorings
(that is, that $i$ and $j$ form a backbone pair) when the chromatic number is
$k$ is $(k/k^2)^\rho=k^{-\rho}$, where $\rho$ is the number of optimal colorings
with $k$ colors. But $\rho$ increases to at least $(k+1)\rho$ when the chromatic
number increases to $k+1$, so the probability we just computed gets divided by
at least $(k+1)^{k\rho}$. We then see that the sudden increases observed in the
number of optimal colorings are probably also inherently related to what happens
to backbones at the same points in the sequence of graphs. This provides a new
perspective on the collapse of backbones at the coloring transitions while at
the same time providing a better understanding of why it happens.

Another well understood combinatorial-growth effect is that, when the chromatic
number is $k$, every optimal coloring is essentially equivalent to $k!-1$
others, each corresponding to a permutation of the colors among the nodes. This
brings us to the third plot of Figure~\ref{poly10}, where the number of optimal
colorings is shown normalized by the factorial of the current chromatic number.
Evidently, all sudden jumps are still there, but they now possess a stronger
significance, because only one optimal coloring is counted out of all colorings
that are equivalent to one another by straightforward permutation of colors
(that is, without implying a different partition of the node set).

But let us return to the chromatic polynomial of $G$. The usefulness of this
polynomial goes beyond the counting of distinct colorings, as it provides the
first link to yet another characterization of the hardness of optimal graph
coloring, now based on the acyclic orientations of $G$. An orientation of $G$ is
an assignment of directions to the edges of $G$; it is acyclic if no directed
cycles are formed, that is, if it is impossible to reach the same node twice by
following edges according to their directions exclusively.

If we let the set of the acyclic orientations of $G$ be denoted by $\Omega(G)$
and consider the number of acyclic orientations of $G$ in terms of what happens
to the graphs $G\backslash e$ and $G/e$ introduced earlier, then we have the
following. Every acyclic orientation of $G\backslash e$ yields either one or two
acyclic orientations for $G$ when $e$ is assigned a direction. The former case
happens when one direction assignment for $e$ forms a directed cycle in $G$
but not the other, the latter when both assignments preserve acyclicity in $G$
(it cannot happen that both assignments form directed cycles because the
orientation of $G\backslash e$ is acyclic). Similarly, every acyclic orientation
of $G/e$ is necessarily one of those acyclic orientations of $G\backslash e$
from which two orientations of $G$ are obtained. Thus,
\begin{equation}
\label{recursion2}
\vert\Omega(G)\vert=\vert\Omega(G\backslash e)\vert+\vert\Omega(G/e)\vert
\end{equation}
for any edge $e$ of $G$.

The recursion in (\ref{recursion2}) is strikingly similar to the one in
(\ref{recursion1}), and this has been shown to give rise to the remarkable
identity
\begin{equation}
\label{nacyclic}
\vert\Omega(G)\vert=(-1)^n\pi(G,-1).
\end{equation}
That is, the number of acyclic orientations of $G$ can be obtained from applying
the chromatic polynomial of $G$ to the negative unit \cite{s73}. Obtaining
(\ref{nacyclic}) and some of its refinements \cite{s99,l01} from the relation
between (\ref{recursion1}) and (\ref{recursion2}) comes as a consequence of the
so-called theory of P-partitions and its order polynomials \cite{s73,v87,s97}.
Of interest to us is that such polynomials quantify the following very useful
relationship between the acyclic orientations of $G$ and its colorings.

Suppose that $G$ can be colored by $k$ colors. Still viewing colors as positive
integers, suppose also that we assign to the edges of $G$ an orientation that
makes every edge point from the node with the higher color to the one with the
lower. This orientation is clearly acyclic and induces no directed path in
$G$ containing more than $k$ nodes. Conversely, suppose we start from an acyclic
orientation of $G$. If $k$ is the number of nodes on the longest directed path
in $G$ according to this orientation, then $G$ can be colored by at most $k$
colors, as follows. We assign color $1$ to the sinks (nodes whose adjacent edges
are all oriented inward), then color $2$ to the sinks that would be formed if
the original sinks were to be removed from $G$, then the lowest available color
to the set of sinks that would appear next, and so on.

\begin{figure}[t]
\centering
\scalebox{0.9}{\includegraphics{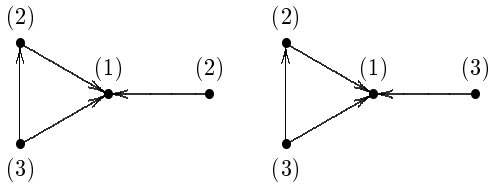}}
\caption{One single acyclic orientation may correspond to more than one
coloring.}
\vspace{0.2in}
\label{1orient2colorings}
\end{figure}

Note, in this process, that starting with a coloring yields a unique acyclic
orientation. The converse is not necessarily true, however: starting with an
acyclic orientation may yield more than one coloring of the nodes of $G$.
Consider, for example, the acyclic orientation shown in
Figure~\ref{1orient2colorings} and the two corresponding colorings shown in
parentheses next to the nodes. What the process indicates, however, is that it
is possible to seek optimal colorings for $G$ by looking for acyclic
orientations of $G$ that are shortest in terms of how many nodes there are in
a longest directed path. Letting $\Omega^-(G)\subseteq\Omega(G)$ be the set of
such orientations, what we have seen is that
\begin{equation}
\vert\Omega^-(G)\vert\le\pi(G,\chi(G)),
\end{equation}
so we may have a tighter characterization of how abundant optimal colorings are
by looking at shortest acyclic orientations instead of the chromatic polynomial
applied to $\chi(G)$.

The formal relationship between $\chi(G)$ and the acyclic orientations of $G$
can be stated as follows \cite{d79}, and strengthens important earlier results
\cite{r67,g68}. For $\omega\in\Omega(G)$, let $P_\omega$ be the set of all
directed paths in $G$ according to $\omega$. For $p\in P_\omega$, let
$\vert p\vert$ indicate the number of nodes in $p$. Then
\begin{equation}
\label{chifromomega}
\chi(G)=\min_{\omega\in\Omega(G)}\max_{p\in P_\omega}\vert p\vert.
\end{equation}

\begin{figure}[t]
\centering
\scalebox{0.9}{\includegraphics{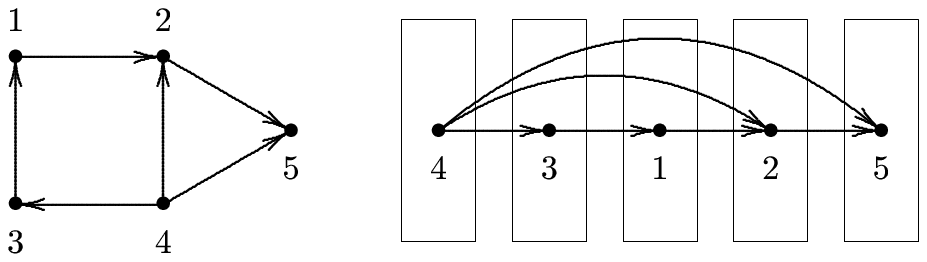}}\\
\vspace{0.1in}
\scalebox{0.9}{\includegraphics{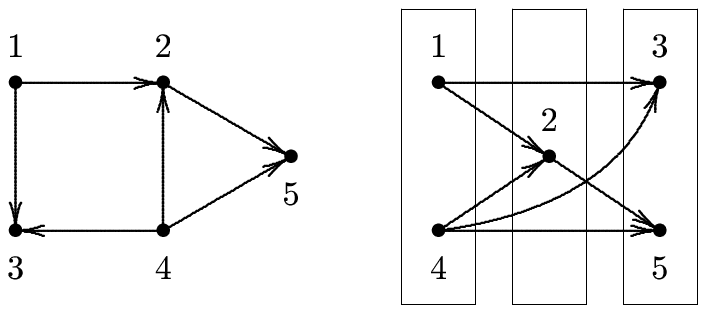}}
\caption{Two acyclic orientations and the corresponding sink decompositions.}
\vspace{0.2in}
\label{length}
\end{figure}

An illustration is given in Figure~\ref{length}, where two acyclic orientations
are shown for the same graph, together with the corresponding partition into
sinks alluded to earlier. This partition is known as the sink decomposition of
the graph according to the acyclic orientation \cite{b76}. In the figure, each
such decomposition is shown with a rightmost box containing the sinks, then
another box to its left containing the sinks that appear if the former sinks
are eliminated, and so on. The set of nodes in each box is normally referred to
as a layer of the sink decomposition. Notice that the number of layers in the
sink decomposition for $\omega$ is precisely $\max_{p\in P_\omega}\vert p\vert$.
In the case of Figure~\ref{length}, the bottommost acyclic orientation has the
smaller sink decomposition and thus corresponds to a better (in this case,
optimal) assignment of colors (a different color to the nodes in each layer).

\begin{figure}[t]
\centering
\scalebox{0.5}{\includegraphics{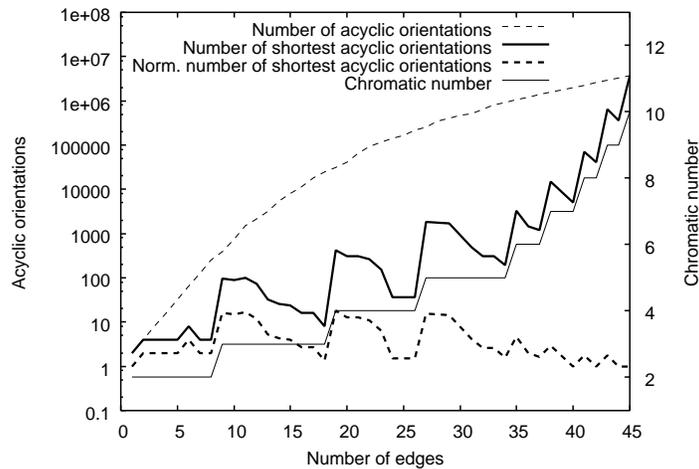}}
\caption{Number of acyclic orientations of each graph in $H_1,\ldots,H_N$ for
$n=10$.}
\vspace{0.2in}
\label{shortest10}
\end{figure}

For $n=10$, Figure~\ref{shortest10} depicts the relationship between the
chromatic number and the number of acyclic orientations of each graph in
$H_1,\ldots,H_N$. The thinner dashed plot in the figure gives the number of
acyclic orientations of each graph. This number has been computed using the
algorithm of \cite{bs99}, which although efficient in several aspects relevant
to the analysis of enumerative algorithms, becomes prohibitive very quickly as
the graph gets larger. The same enumeration process has been used to record the
number of acyclic orientations that are shortest, that is, those whose sink
decompositions have as many layers as the graph's chromatic number. This small
addition to the algorithm employs straightforward depth-first search
\cite{clrs01}, and the results are shown as the thicker solid plot of the
figure.

\begin{figure}[t]
\centering
\scalebox{0.5}{\includegraphics{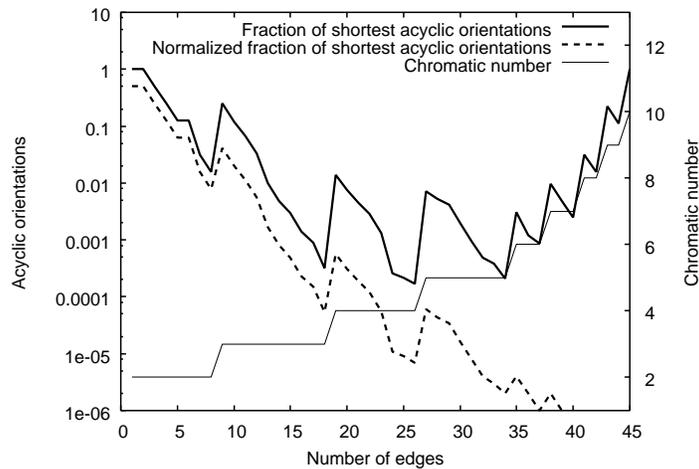}}
\caption{Relative occurrence of shortest acyclic orientations of each graph in
$H_1,\ldots,H_N$ for $n=10$.}
\vspace{0.2in}
\label{compcolor10}
\end{figure}

Remarkably, an effect very similar to the one observed in Figure~\ref{poly10} is
seen to occur now as well: the number of shortest acyclic orientations increases
sharply whenever the chromatic number increases, and subsequently goes down
until immediately before the next increase. These are also absolute data, but
now it is obvious how to make them relative: we simply observe the percentage
of all acyclic orientations that are shortest. This is shown in
Figure~\ref{compcolor10}, which confirms that, also in relative terms, shortest
acyclic orientations become significantly more abundant right after an increase
in the chromatic number, becoming increasingly rarer from there onward until
the next increase.

This indicates that, in a sense, coloring a graph optimally becomes easier
immediately after an increase in the chromatic number, and incrementally harder
until the next transition is reached. Behind this statement is the intuitive
feeling that coloring algorithms should in general fare better when optimal
solutions are more abundant, be such abundance assessed as some indicator of how
many optimal colorings there are or as the fraction
$\vert\Omega^-(G)\vert/\vert\Omega(G)\vert$ of shortest acyclic orientations. Of
course, this intuition calls for experimental support, but notice that at least
the most na\"\i ve of all random approaches---a series of Bernoulli trials
inside $\Omega(G)$ until the first member of $\Omega^-(G)$ is found---is certain
to benefit from such abundance of shortest acyclic orientations, as clearly the
expected time for its convergence is $\vert\Omega(G)\vert/\vert\Omega^-(G)\vert$
\cite{f68}. More serious methods with the potential to benefit from the relative
abundance of optimal acyclic orientations exist \cite{ban03}, though, and a
systematic effort to assess their capabilities on sequences of random graphs is
under way.

The number (or fraction) of optimal acyclic orientations is, in principle,
also subject to the same concerns with the disguising action of obvious
combinatorial growth that we expressed earlier. To see what happens when we
apply the same normalization by the factorial of the current chromatic number,
we have in each of Figures~\ref{shortest10} and \ref{compcolor10} a plot with
the results (the one in thick dashes). The pronounced jumps are still present,
but this normalization is only meaningful as inherited from the relationship
between optimal colorings and shortest acyclic orientations.

A better normalization may exist and we would like to digress on this
possibility briefly, although several problems related to it are still open.
Suppose we take an acyclic orientation and turn all its sinks into sources
(nodes whose adjacent edges are all oriented outward). This necessarily yields
another acyclic orientation, and the continual repetition of the process must
eventually lead to a period of orientations. This attractor dynamics has several
interesting properties \cite{bg89,b00}; in our context, the most crucial
property is that the number of layers in the graph's sink decomposition is
continually nonincreasing along the process of obtaining new acyclic
orientations by turning sinks into sources. So the period at the core of each
attractor comprises orientations that yield sink decompositions all with the
same number of layers, this number being also no larger than that resulting from
any other orientation in the same attractor. This means that, in addition to the
chromatic indicator originally observed when this attractor dynamics was first
analyzed (the graph's interleaved multichromatic number, cf.\ \cite{b02}), each
attractor is related to finding the chromatic number as well, since finding a
period whose orientations are shortest over all attractors immediately yields
the chromatic number. A period, in summary, provides a means of expressing the
equivalence of several acyclic orientations and may become suitable for the
normalization we need if only more knowledge can be obtained on it.
Unfortunately, we thus far lack this necessary additional knowledge.

\begin{figure}[t]
\centering
\scalebox{0.9}{\includegraphics{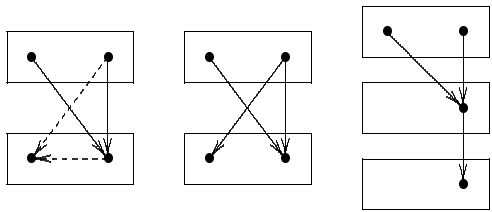}}
\caption{Two choices when adding a new directed edge and the resulting sink
decompositions.}
\vspace{0.2in}
\label{morelayers}
\end{figure}

To finalize, we comment on yet another interesting insight that can be gained
from considering the relationship between a graph's colorings and its acyclic
orientations. Suppose that $\omega$ is an acyclic orientation of $G$, and
consider the random addition of an edge to $G$ between two nodes not currently
connected. If we look at $\omega$ from the perspective of its sink
decomposition, then let its layers be numbered $1$ through $L$, starting at the
layer that contains the sinks and onward. As illustrated in
Figure~\ref{morelayers}, in some cases the addition of the new edge will
preserve the sink decomposition, while in others it will not. These two
possibilities are shown in the leftmost sink decomposition in the figure as two
dashed directed edges. The addition of one of them preserves the sink
decomposition (shown in the middle sink decomposition of the figure); the
addition of the other, which connects two nodes in the same layer, forces the
sink decomposition to acquire another layer, as shown in the rightmost sink
decomposition of the figure.

Let us then assume that, if the new edge is added between layers, then it is
oriented in the direction of the lower-numbered layer. Let also the $\ell$th
layer have $s_\ell$ nodes, so that $n=\sum_{\ell=1}^Ls_\ell$. The number of
node pairs not yet connected by an edge is $N-m$; of these, the ones that have
nodes in different layers amount to
$\sum_{\ell=1}^{L-1}\sum_{\ell'=\ell+1}^Ls_\ell s_{\ell'}-m$.
If we let $q$ be the probability that adding the new edge does not disturb the
sink decomposition, then $q$ is the probability that the edge is added between
layers. That is,
\begin{equation}
\label{prob}
q=\frac
{\sum_{\ell=1}^{L-1}\sum_{\ell'=\ell+1}^Ls_\ell s_{\ell'}-m}
{N-m}.
\end{equation}

By (\ref{prob}), the addition of an edge between layers (this happens with
probability $q$) causes $q$ to decrease, as the only change in the formula is
the concomitant subtraction of $1$ off both the numerator and the denominator. 
When the edge is added between nodes of the same layer (with probability $1-q$),
the double summation in the numerator of (\ref{prob}) may either remain the same
or vary. If it remains the same or decreases, then $q$, as before, decreases. In
order to verify what happens otherwise, let $S$ denote the double summation.
Then $q$ is seen to vary by at least
\begin{equation}
\frac{S+1-m-1}{N-m-1}-\frac{S-m}{N-m}=\frac{S-m}{(N-m)^2-(N-m)}\ge 0,
\end{equation}
so $q$ increases unless $S=m$, in which case it was zero to begin with and
remains zero.

If we now consider the number of edges that need to be randomly added to $G$ so
that $q$ once again assumes the value it currently has, say $\epsilon$, it is
easy to see from (\ref{prob}) that this number is given by
$\delta/(1-\epsilon)$, where $\delta$ is the increase that $q$ incurs along the
way. The value of $\delta$ is very hard to quantify, but it seems reasonable to
assume, at least for the sake of the argument, that it gets smaller as the graph
gets denser (i.e., acquires more edges). In this case, the number of random
edge additions needed for $q$ to return to the value $\epsilon$ is ever smaller
as the process unfolds. Overall, what we witness is a process that resembles the
increase in $\chi(G)$ as $G$ becomes denser, slow at first but increasingly
rapid as the graph's density gets higher. We hope to obtain a better
characterization of this resemblance as further research adds detail to the
picture. If we succeed, it may be possible to obtain a generic prescription for
determining all the points at which the chromatic number increases along the
sequence, thus adding to what is already known \cite{bp01,mpwz02}.

\section{Independent sets}\label{mis}

\begin{figure}[t]
\centering
\scalebox{0.9}{\includegraphics{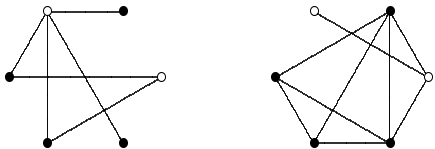}}
\caption{A maximum independent set (filled circles in the graph on the left) and
the corresponding maximum clique in the complementary graph (filled circles in
the graph on the right).}
\vspace{0.2in}
\label{maxclique}
\end{figure}

Let $\alpha(G)$ denote the independence number of $G$. We have used
public-domain code \cite{dfmax.c-url} based on \cite{cp90} to find $\alpha(H_s)$
for each graph in $H_1,\ldots,H_N$ with $n=75$. What this code finds is not a
maximum independent set directly, but rather a maximum clique in the graph that
is complementary to the graph of interest (i.e., has an edge joining two
distinct nodes if and only if the graph of interest does not). A clique is a
subgraph whose nodes are all connected to one another, so the correspondence to
independent sets should be clear. This is illustrated in Figure~\ref{maxclique}.

\begin{figure}[t]
\centering
\scalebox{0.5}{\includegraphics{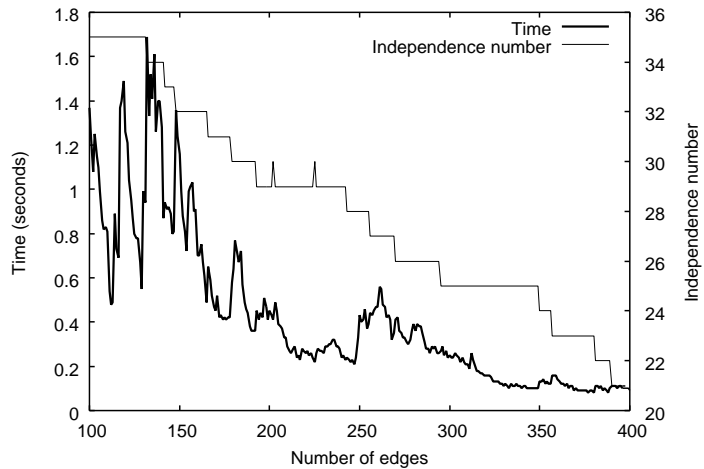}}
\caption{Time to find the independence number of some of the graphs in
$H_1,\ldots,H_N$ for $n=75$.}
\vspace{0.2in}
\label{dfmax}
\end{figure}

Figure~\ref{dfmax} has two plots for $n=75$, one to indicate the time to find
the graph's independence number, the other to indicate how this number evolves
as the graph becomes denser. We show data for a certain density interval only.
To the left of what is shown the graphs have isolated nodes and the behavior
is uncharacteristic, while to the right times become too small to be indicative
of any particularly interesting behavior. But inside the density interval used
in the figure the independence number goes down nearly steadily,\footnote{The
two increases in the independence number back to $30$ from $29$ for $s=202$ and
$s=225$ can only be explained by the existence of two last isolated nodes in
$G_{201}$, one of which gets incorporated into $H_{202}$, the other into
$H_{225}$. That a node should remain isolated through $s=224$ for $n=75$ is
unlikely but entirely conceivable. In fact, the probability of such an event,
given approximately by $e^{-2s/n}$ (cf.\ (\ref{isolated})), is $0.0025$.}
and at some of these downward transitions there appears to be a considerable
decrease in the time for optimal solution right after the decrease in the
independence number. Although this evidence is less ubiquitous than in the case
of graph coloring, and for this reason certainly less compelling, this study
seems to be the first one to address the issue of phase transitions related to
finding a graph's independence number and for this reason we investigate the
matter further. In this section, we introduce some combinatorial arguments that
appear to relate to these phenomena.

As for graph coloring, our initial step is to assess the abundance of maximum
independent sets for a given graph $G$, that is, independent sets of size
$\alpha(G)$. Unlike the case of graph coloring, however, it is as yet unknown
how to count this number exactly. This remains true even if we settle for
maximal (as opposed to maximum) independent sets, that is, independent sets that
cannot be enlarged without losing the independence property, although in this
case increasingly better upper bounds on the desired number have been discovered
recently (cf.\ \cite{n02} and the references therein).

But at this point it helps to recall that the sequence of graphs
$G_1,\ldots,G_N$ is randomly generated, each graph being drawn uniformly from
the set of graphs having the same number of nodes and edges. We may therefore
attempt to assess the number of maximal independent sets of a given size in
each graph by computing the expected value of this number, which is in fact a
random quantity. In order to accomplish this more easily, it is helpful to
resort to the model of random graphs in which an edge exists between two nodes
with constant probability, say $p$, independently of the nodes. Although this
is not the model under which our graphs were generated, away from limiting
situations it is safe to assume that the two models are equivalent to each
other with $p=m/N$ for $G(n,m)$ \cite{b01}.

Let $\xi(G,k)$ denote the expected number of maximal independent sets of size
$k$ in $G$. The number of candidate sets is $n\choose k$, and the probability
that each one is an independent set is $(1-p)^{k(k-1)/2}$. If a candidate is an
independent set, then the probability that it is maximal is the probability that
each of the remaining $n-k$ nodes is connected to at least one of its $k$ nodes,
that is, $[1-(1-p)^k]^{n-k}$. We then get
\begin{equation}
\label{expectedmis}
\xi(G,k)={n\choose k}
\left(1-\frac{m}{N}\right)^\frac{k(k-1)}{2}
\left[1-\left(1-\frac{m}{N}\right)^k\right]^{n-k}.
\end{equation}
Now let $G$ be a random graph generated in compliance with the edge density
given by $p$, and let the value of $\alpha(G)$ be known. Letting $k=\alpha(G)$
in (\ref{expectedmis}) yields the expected number of maximal independent sets in
similar random graphs, whose size is, at least for one of them, the size of its
maximum independent set.

\begin{figure}[t]
\centering
\scalebox{0.5}{\includegraphics{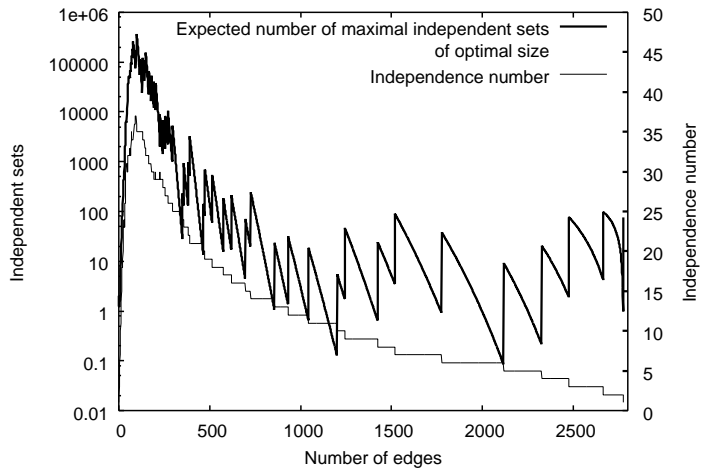}}
\caption{Expected number of size-$\alpha(H_s)$ maximal independent sets
of each graph in $H_1,\ldots,H_N$ for $n=75$.}
\vspace{0.2in}
\label{expected75}
\end{figure}

For $n=75$, we show the evolution of this number for the sequence
$H_1,\ldots,H_N$ in Figure~\ref{expected75}, along with the evolution of
the independence number. Interestingly, every transition of the independence
number to a smaller value causes the expected number of maximal independent sets
of size $\alpha(H_s)$ to increase markedly. From there onward, this number
decreases until the next similar transition occurs. This is one first
indication, albeit imprecise, that the abundance of such maximal sets may be
related to the increased ease with which the graph's independence number can
sometimes be found immediately after a decrease in the independence number.
However, as in our analysis of the chromatic polynomial in
Section~\ref{coloring}, it is not clear how to proceed and characterize the
relationship more effectively. Not only this, but it is still possible, as in
the case of graph coloring, that some underlying inherent equivalence among
maximal independent sets exists that would indicate a way of normalizing the
data shown in the figure. We still do not know how that can be achieved.

\begin{figure}[t]
\centering
\scalebox{0.9}{\includegraphics{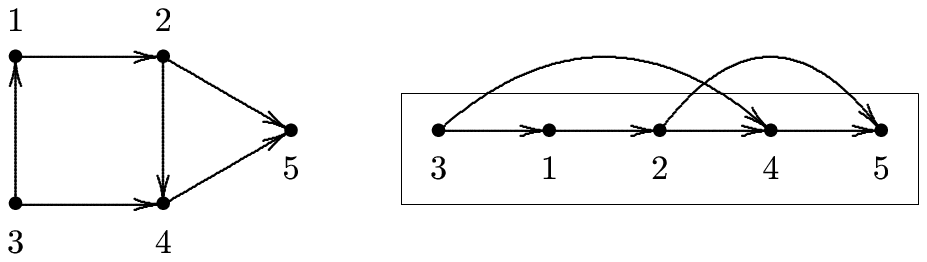}}\\
\vspace{0.1in}
\scalebox{0.9}{\includegraphics{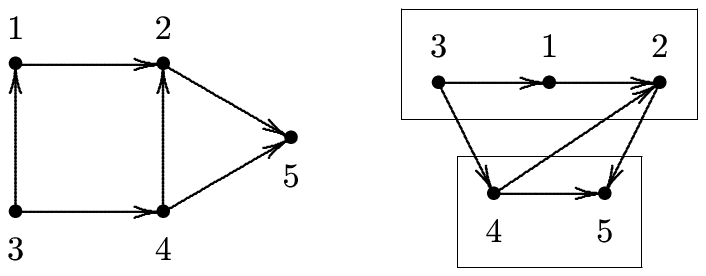}}
\caption{Two acyclic orientations and the corresponding minimum chain
decompositions.}
\vspace{0.2in}
\label{width}
\end{figure}

Once again, though, a deep relationship exists between a graph's independent
sets and its acyclic orientations. It is in this case both more complex and more
subtle than in the case of graph coloring, so we describe it with the aid of an
illustration right from the start. First consider Figure~\ref{width}, where two
acyclic orientations of the same graph are shown, each one alongside what is
known as a chain decomposition of the graph according to it. A chain
decomposition of a graph according to an acyclic orientation is a partition of
the graph's node set such that the nodes in each set of the partition are
arranged by the acyclic orientation as a single chain of nodes (a directed
path). In the case of Figure~\ref{width}, each partition is displayed with the
aid of boxes to enclose the nodes that go in each set. Furthermore, each of the
chain decompositions shown in the figure employs the minimum number of chains
for the corresponding acyclic orientation. The topmost acyclic orientation
admits a chain decomposition comprising one single chain, while for the other no
chain decomposition exists with fewer than two chains.

\begin{figure}[t]
\centering
\scalebox{0.9}{\includegraphics{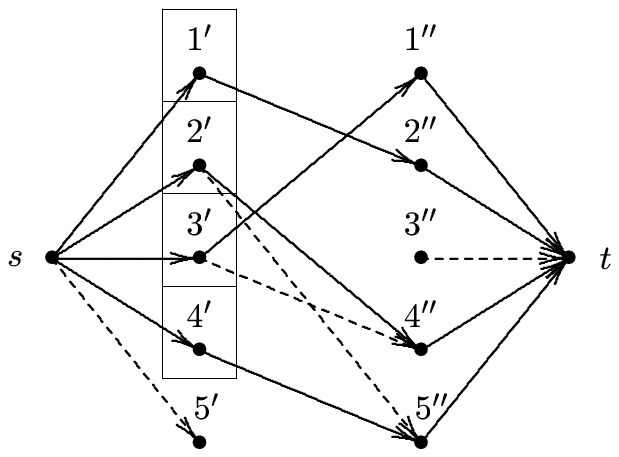}}\\
\vspace{0.1in}
\scalebox{0.9}{\includegraphics{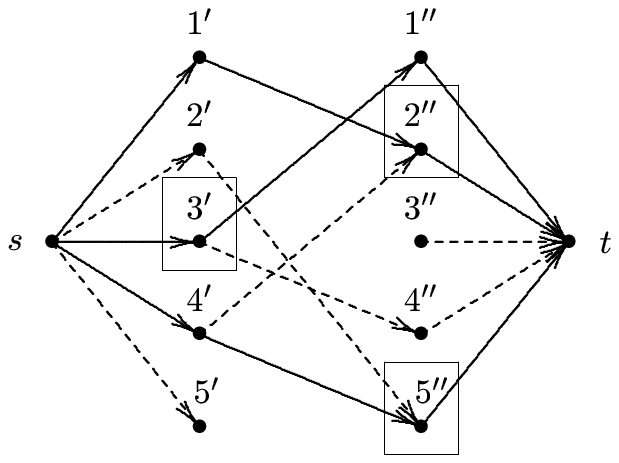}}
\caption{Flow networks associated with Figure~\ref{width}.}
\vspace{0.2in}
\label{flow}
\end{figure}

Understanding how chain decompositions and independent sets are related
involves several technicalities that we will not discuss formally but illustrate
through examples instead. Figure~\ref{flow} contains two flow networks, that is,
two directed graphs with two distinguished nodes, $s$ and $t$, on which we
consider the maximum flow from $s$ to $t$ subject to certain edge capacities
(unit for all edges that emanate from $s$ or arrive at $t$, infinite for all
others). Each flow network corresponds to one of the acyclic orientations of
Figure~\ref{width} and is constructed from that orientation as follows. For
each node $i$ in $G$, two nodes, $i'$ and $i''$, are added to the flow network, along with a directed edge from $s$ to $i'$ and one from $i''$ to $t$.
If an edge exists between nodes $i$ and $j$ and the acyclic orientation directs
that edge from $i$ to $j$, then the flow network contains an edge directed from
$i'$ to $j''$.

The maximum flow from $s$ to $t$ in each of the networks of Figure~\ref{flow} is
indicated by solid edges (edges that carry unit flow) and dashed edges (edges
that carry zero flow). In general, the following interesting property holds for
maximum flows in such networks \cite{amo93}. Of the solid edges, those whose
removal disconnects $s$ from $t$ minimally (the minimum cut) induce a node cover
in $G$, that is, a set of nodes that touches every edge. In Figure~\ref{flow},
an edge in the minimum cut either leads from $s$ to a node enclosed in a box or
from such a node to $t$. Nodes thus represented correspond to the node sets
$\{1,2,3,4\}$ and $\{2,3,5\}$ in $G$, each set a node cover. Note that, by
definition, the complements of these sets with respect to the node set of $G$
are necessarily independent sets of $G$---in the figure, these are the sets
$\{5\}$ and $\{1,4\}$, each of whose nodes corresponds to one of the chains in
Figure~\ref{width}.

Formally, the relationship between $\alpha(G)$ and the acyclic orientations of
$G$ is as given next \cite{d79}, and brings sharper focus to the classic result
established by Dilworth's theorem \cite{d50}. For $\omega\in\Omega(G)$, let
$D_\omega$ be the set of all chain decompositions of $G$ according to $\omega$.
For $d\in D_\omega$, let $\vert d\vert$ be the number of chains in $d$. Then
\begin{equation}
\label{alphafromomega}
\alpha(G)=\max_{\omega\in\Omega(G)}\min_{d\in D_\omega}\vert d\vert.
\end{equation}
As one readily notices, (\ref{alphafromomega}) is a sort of dual of
(\ref{chifromomega}): it expresses $\alpha(G)$ as the number of chains in the
minimum chain decomposition that has the most chains over $\Omega(G)$.
Henceforth, we let $\Omega^+(G)\subseteq\Omega(G)$ denote the set of such
widest acyclic orientations, understood as those acyclic orientations that
achieve the minimum chain decomposition that is maximum over $\Omega(G)$.

It is now instructive to return briefly to Figure~\ref{width}. For each
acyclic orientation $\omega$ displayed in the figure, the number of chains
in the corresponding chain decomposition is $\min_{d\in D_\omega}\vert d\vert$.
Of the two acyclic orientations, the bottommost achieves the maximum of this
quantity over $\Omega(G)$.

We now investigate the use of $\vert\Omega^+(G)\vert$ as an indicator of how
abundant the widest acyclic orientations of $G$ are and how this relates to the
hardness of finding $\alpha(G)$. This time, our procedure to enumerate all the
acyclic orientations of a graph has been coupled with a public-domain code
\cite{hipr.tar-url} that implements the algorithm of \cite{cg97} to find the
maximum flow in a network (and also the minimum cut, as a by-product).

\begin{figure}[t]
\centering
\scalebox{0.5}{\includegraphics{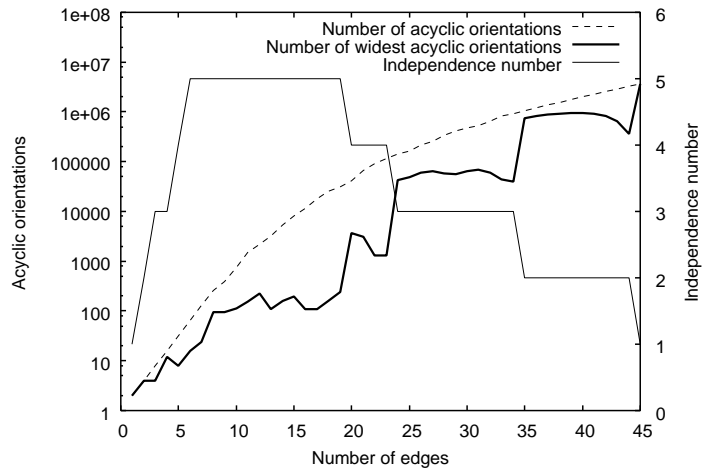}}
\caption{Number of acyclic orientations of each graph in $H_1,\ldots,H_N$ for
$n=10$.}
\vspace{0.2in}
\label{widest10}
\end{figure}

The results are shown in Figure~\ref{widest10}, in which the number of all
acyclic orientations of each graph is plotted alongside its independence number
and the number of acyclic orientations that are widest (those whose chain
decomposition into the fewest possible chains requires $\alpha(H_s)$ chains).
The figure indicates that the same sharp transitions that appear in
Figure~\ref{expected75} are also observed for the acyclic orientations: every
downward transition that the independence number undergoes is accompanied by a
sharp increase in the number of widest acyclic orientations.

\begin{figure}[t]
\centering
\scalebox{0.5}{\includegraphics{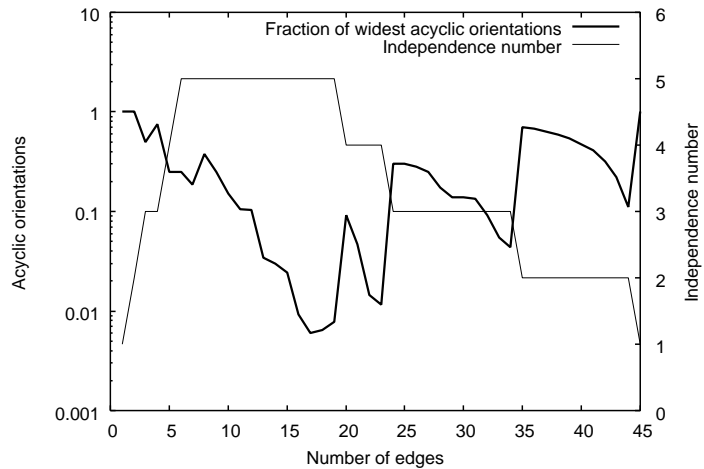}}
\caption{Relative occurrence of widest acyclic orientations of each graph in
$H_1,\ldots,H_N$ for $n=10$.}
\vspace{0.2in}
\label{compmis10}
\end{figure}

When we assess such increases with respect to the set of all the acyclic
orientations of each graph, we obtain what is shown in Figure~\ref{compmis10}.
Clearly, every decrease in the independence number corresponds to a marked
increase in the fraction of acyclic orientations that are widest. We hope to
eventually be able to conclude that such increases are correlated with the
time it takes to find a graph's maximum independent set. As we mentioned in
Section~\ref{coloring}, obviously the unreasonable approach of random trials
does benefit from an elevated $\vert\Omega^+(G)\vert/\vert\Omega(G)\vert$ ratio,
but we believe this may also be the case for heuristics that exploit the role of
acyclic orientations directly \cite{bc03}.

\section{Conclusions}\label{concl}

We have investigated the NP-hard problems of coloring the nodes of a graph
optimally and of finding a maximum independent set in a graph. For each of
these problems, we first displayed empirical evidence that, as the number of
edges in the graph is increased by the random addition of one edge at a time,
significant variations take place in the time to solve the problem optimally.
Often these variations coincide with the transition of the value of the optimal
solution to a new level, a higher one for graph coloring, a lower one for
independent sets.

For graph coloring, we have shown that the upward transitions in the chromatic
number along the sequence of increasingly denser graphs coincide with sharp
increases in the abundance of distinct colorings employing the optimal number
of colors. More importantly, we have shown that the same phenomenon takes place
when we consider the shortest acyclic orientations of the graphs, that is, those
orientations whose sink decompositions have as many layers as the graphs'
chromatic numbers. In this case, sharp increases are observed in the ratio
$\vert\Omega^-(H_s)\vert/\vert\Omega(H_s)\vert$.

Our conclusions for the maximum independent set problem are similar. The
downward transitions in the independence number that occur along the sequence of
graphs coincide with sharp increases in the expected number of maximal
independent sets of certain sizes. As for the acyclic orientations of the
graphs, we have given evidence that the same type of phenomenon takes place
regarding the widest acyclic orientations of the graphs, that is, those whose
minimum chain decompositions have as many chains as the graphs' independence
numbers. In this case, sharp increases take place in the ratio
$\vert\Omega^+(H_s)\vert/\vert\Omega(H_s)\vert$.

We find that the two special subsets of $\Omega(G)$, $\Omega^-(G)$ and
$\Omega^+(G)$, underlie an interpretation of the phase transitions observed in
optimal graph coloring and optimal independent sets that is full of
combinatorial meaning. This meaning is directly related to the structure of the
graphs involved and that of its acyclic orientations. In addition, it provides
a direct interpretation of the hardness of the problems as given by the relative
abundance of the acyclic orientations that give the optima.

Except for those shown in Figures~\ref{trick}, \ref{dfmax}, and
\ref{expected75}, all our empirical results have been given for $n=10$ only. As
we indicated earlier, the reason for this has been the severe combinatorial
explosion that occurs when finding a graph's chromatic polynomial or its set of
acyclic orientations. For $n=15,20$, so far these computations could only be
carried out through about $40$ edges, yielding results that fully support the
conclusions we have drawn along the paper. But, instead of presenting such
partial results, we opted for giving the reader data on the full evolution
through the densest graphs for $n=10$.

We think the empirical evidence we have provided is only the beginning of a
deeper investigation of how the abundance of certain acyclic orientations
affects the hardness of optimal graph coloring and of finding maximum
independent sets. Given the aforementioned computational difficulties, the most
promising avenue for continuation seems to be the search for additional analytic
properties that can explain the findings we have described. Perhaps the
evolution of the probability in (\ref{prob}) superficially described at the end
of Section~\ref{coloring}, and of an analogous indicator in the case of
independent sets, should be investigated more deeply as a first step.

\subsection*{Acknowledgments}

The authors acknowledge partial support from CNPq, CAPES, the PRONEX initiative
of Brazil's MCT under contract 41.96.0857.00, and a FAPERJ BBP grant.


\end{document}